\documentclass[11pt]{article}
\usepackage{vietnam,epsfig}

\def\be{\begin{equation}}
\def\ee{\end{equation}}
\def\bea{\begin{eqnarray}}
\def\eea{\end{eqnarray}}

\begin{document}
\vspace*{4cm}
\title{A NEW DARK MATTER CANDIDATE IN LOW-TENSION BRANE-WORLDS}

\author{ J.A.R. CEMBRANOS$^{1,\;3}$, A. DOBADO$^{2,\;3}$ AND A.L. MAROTO$^3$ }

\address{$^1$ Department of Physics and Astronomy,
 University of California, Irvine, CA 92697, USA\\
$^2$ Stanford Linear Accelerator Center, 2575 Sand Hill Road,
Menlo Park, CA 94025, USA \\
$^3$ Departamento de  F\'{\i}sica Te\'orica,
 Universidad Complutense de
  Madrid, 28040 Madrid, Spain}

\maketitle\abstracts{
Brane world theories contain additional degrees of freedom related
to the geometry of the extra dimension space which can be
understood as brane oscillations (branons). In the case where the
fundamental gravitational scale is much larger than the brane
tension scale, these branons are the only extra degrees of freedom
at low energies coming from the extra dimensions. Branons are
generically stable, weakly interacting and massive. They could be
produced in the next generation colliders and at the same time
they are natural WIMP like candidates for dark matter.}
\section{The branon field}
The Brane World (BW) models were proposed at the final of the
past century, with the interesting possibility to have
observable effects at the scale of present or near
experiments \cite{ADD}. The main idea that defines the
BW scenario is that the Standard Model (SM) particles are
restricted to a three-dimensional hypersurface or 3-brane,
whereas the gravitons can propagate along the whole bulk space.

Since rigid objects do not exist in relativistic theories, it
is clear that brane fluctuations must play an important role in this framework \cite{DoMa}. This fact turns out to be
particularly true when the brane tension scale $f$
($\tau=f^4$ being the brane tension) is much smaller than the
$D$ dimensional or
fundamental gravitational scale $M_D$, i.e. $f<<M_D$. In this
case the only relevant low-energy modes of the BW scenarios are
the SM particles and branons which are the quantized brane
 oscillations. Indeed branons can be understood as the (pseudo-)Goldstone
 bosons corresponding to the spontaneous breaking of translational
 invariance in the bulk space produced by the presence of the brane.

The branon properties allow to solve some of the problems
of the brane-world scenarios such as the
divergent virtual contributions from the Kaluza-Klein tower at the
tree level or non-unitarity of the graviton production
cross-sections \cite{GB}.
The SM-branon low-energy effective Lagrangian reads
\cite{DoMa,BSky,ACDM}:
\begin{eqnarray}
{\mathcal L}_{Br}&=&
\frac{1}{2}g^{\mu\nu}\partial_{\mu}\pi^\alpha
\partial_{\nu}\pi^\alpha-\frac{1}{2}M^2\pi^\alpha\pi^\alpha
+\frac{1}{8f^4}(4\partial_{\mu}\pi^\alpha
\partial_{\nu}\pi^\alpha-M^2\pi^\alpha\pi^\alpha g_{\mu\nu})
T^{\mu\nu}_{SM}\label{lag}
\end{eqnarray}
We see that branons interact by pairs with the SM
energy-momentum tensor. This means that they are stable particles.
On the other hand, their couplings are suppressed by the
brane tension $f^4$, i.e. they are weakly interacting. These features
make them natural dark matter \cite{CDM,M} candidates (see \cite{MaRa}
for updated reviews on cosmology and dark matter).
\section{Brane World dark matter}
When the branon annihilation rate, $\Gamma=n_{eq}\langle\sigma_A v
\rangle$,
equals the universe expansion rate $H$, the branon abundance
freezes out relative to the entropy density. This happens at the
so called freeze-out temperature $T_f=M/x_f$. We have computed this
relic branon abundance in two cases: either relativistic branons at
freeze-out (hot-warm) or non-relativistic (cold), and assuming that
the evolution of the universe is standard for $T<f$ (see Fig. 1).
\begin{figure}[h]
\centerline{\epsfxsize=12cm\epsfbox{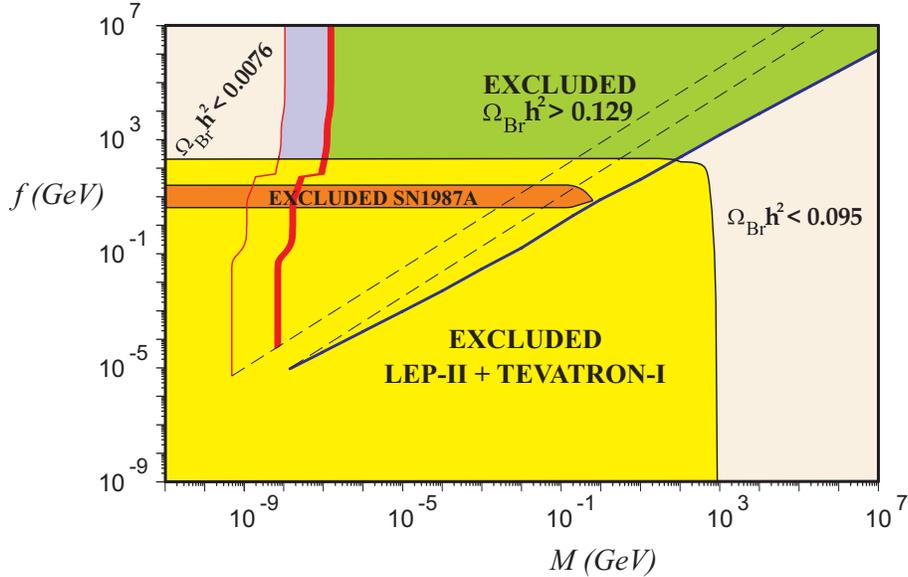}}
\caption {\footnotesize Relic abundance in the $f-M$ plane for a model with one branon of
mass: $M$. The two lines on the left correspond to the $\Omega_{Br}h^2=0.0076$ and
$\Omega_{Br}h^2=0.129 - 0.095$ curves for hot-warm relics, whereas the right line
corresponds to the latter limits for cold relics.
The lower area is excluded by single-photon processes at LEP-II together
with monojet signal at Tevatron-I. The upper area is also excluded by
cosmological branon overproduction. The astrophysical constraints are less
restrictive and they mainly come from supernova cooling by branon emission.}
\end{figure}
On the other hand, if branons make up the galactic halo, they could be detected by direct search
experiments from the energy transfer in elastic collisions with nuclei of a
suitable target. For the allowed parameter region in Fig. 1, branons cannot
be detected by present experiments such as DAMA, ZEPLIN 1 or EDELWEISS.
However, they could be observed by future detectors such as CRESST II, CDMS or
GENIUS.

Branons could also be detected indirectly: their annihilations in the galactic
halo can give rise to pairs of photons or $e^+ e^-$ which could be detected by
$\gamma$-ray telescopes such as MAGIC or GLAST or antimatter detectors
(see \cite{CDM} for an estimation of positron and photon fluxes from
branon annihilation in AMS).
Annihilation of branons trapped in the center of the sun or the earth can
give rise to high-energy neutrinos which could be detectable by high-energy
neutrino telescopes such as AMANDA, IceCube or ANTARES.
\section{Branon signals in colliders}
The dark matter searches complement those in high-energy particle colliders. The branon signals depend on their number
$N$, the brane tension scale $f$, and their masses $M$. From the effective action given in the Equation (1), one can calculate the relevant cross-sections for different branon searches. The single photon channel and the monojet production are the more interesting ones.
\begin{figure}[h]
\begin{center}
\resizebox{7.5cm}{!}{\includegraphics{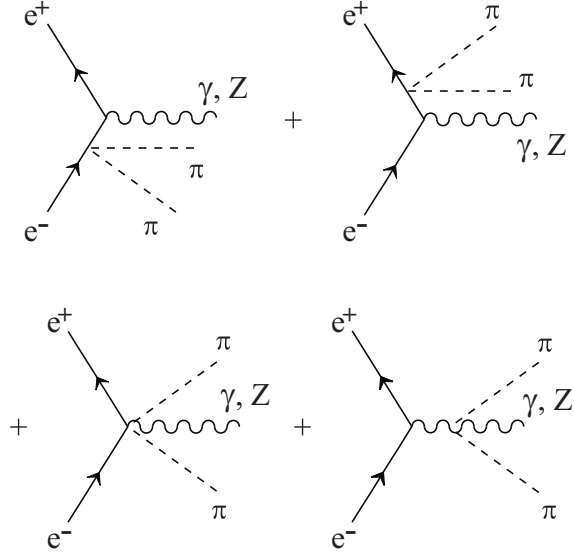}}
\caption {\footnotesize Relevant Feynman diagrams for the branon
contribution to the
single Z and the single photon channel.} \label{LC}
\end{center}
\end{figure}
The main results in relation with these analysis are presented in Table 1, where one can find not only the present restrictions coming from HERA, Tevatron and LEP-II but also the prospects for future colliders like ILC, LHC or CLIC \cite{ACDM,L3}.
\begin{table}[h]
\centering \small{
\begin{tabular}{|c|ccccc|}
\hline\hline Experiment
&
$\sqrt{s}$(TeV)& ${\mathcal
L}$(pb$^{-1}$)&$\sigma_0$(GeV$^{-2}$)&$f_0$(GeV)&$M_0$(GeV)\\
\hline
%
%
HERA$^{\,1}
$& 0.3 & 110 & $7.0\, 10^{-7}$ & 16 & 152
\\
Tevatron-I$^{\,1}
$& 1.8 & 78 &  $6.3\, 10^{-10}$ & 157 & 822
\\
Tevatron-I$^{\,2}
$ & 1.8 & 87 & $1.3\,10^{-10}$ & 148 & 872
\\
LEP-II$^{\,2}
$& 0.2 & 600 & $3.3\, 10^{-11}$ & 180 & 103
\\
\hline
Tevatron-II$^{\,1}
$& 2.0 & $10^3$ & $3.2\, 10^{-10}$ & 256 & 902
\\
Tevatron-II$^{\,2}
$& 2.0 & $10^3$ &  $7.0\, 10^{-11}$ & 240 & 952
\\
ILC$^{\,2}
$& 0.5 & $2\, 10^5$ & $1.5\, 10^{-11}$ & 400 & 250
\\
LHC$^{\,1}
$& 14 & $10^5$ & $1.8\, 10^{-11}$ & 1075 & 6481
\\
LHC$^{\,2}
$& 14 & $10^5$ &  $3.8\,10^{-12}$ & 797 & 6781
\\
CLIC$^{\,2}
$& 5 & $10^6$ & $6.6\, 10^{-12}$ & 2640 & 2500
\\
\hline\hline
\end{tabular}
} \caption{\footnotesize{Summary of the main analysis related to collider experiments. All the results are
performed at the $95\;\%$ c.l. Two different channels have been studied: the
one marked with an upper index $^1\,$ is related to monojet
production, whereas the single photon is labelled with an upper
index $^2\,$. The table contains a total of seven experiments: HERA, LEP-II, the I and II Tevatron runs, ILC, LHC and CLIC. Obviously the data corresponding to the four last experiments are estimations, whereas the first three analysis have been performed with real data.
$\sqrt{s}$ is the center of mass energy associated to the total
process; ${\mathcal L}$ is the total integrated luminosity;
$\sigma_0$ is the estimation for the cross
section sensitivity limit; $f_0$, the bound in the brane tension scale for one massless branon ($N=1$) and $M_0$ the limit on the branon mass for $f=0$.}}
\label{tabHad}
\end{table}

On the other hand, the branon radiative correction on the SM
phenomenology could be important. The one loop
calculation generates higher-dimensional operators
involving SM fields, suppressed by powers of the brane
tension scale \cite{CrSt,GS}. One of the most relevant
contributions of virtual branons to the phenomenology of the
SM particles could be the effects on four-fermion interactions.
For a generic four-fermion process, the branons induce a new
effective vertex as the Figure 3 shows.
\begin{center}
\begin{figure}[h]
\begin{tabular}{c c c}
\begin{tabular}{c c c}
$\psi_a(p_2)$& &$\psi_b(p_4)$\\
 &\epsfysize=2.7 cm\epsfbox{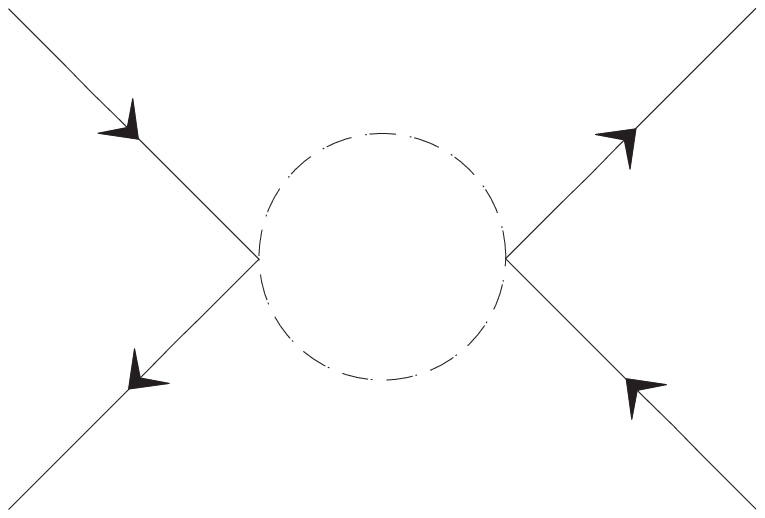}& \\
 $\bar\psi_a(p_1)$& &$\bar\psi_b(p_3)$
\end{tabular}
 &
$\Rightarrow$
&
\begin{tabular}{c c c}
$\psi_a(p_2)$& &$\psi_b(p_4)$\\
 &\epsfysize=2.7 cm\epsfbox{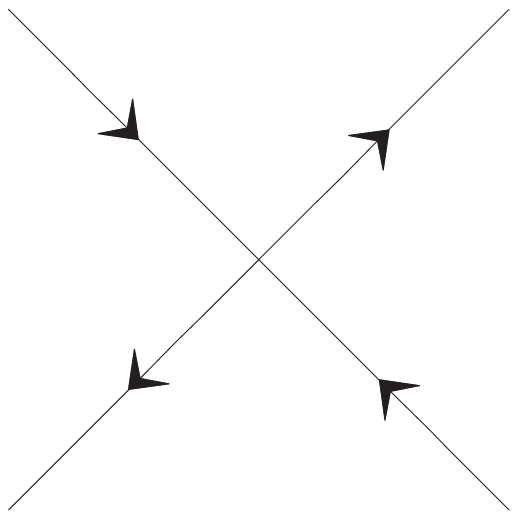}& \\
 $\bar\psi_a(p_1)$& &$\bar\psi_b(p_3)$
\end{tabular}
\end{tabular}
\caption {\footnotesize Four-fermion vertex induced by branon radiative corrections.}
\end{figure}
\end{center}

This kind of signals can be studied in the above mentioned
colliders.
The branons also can generate an anomalous magnetic
moment for each charged fermion. However this effect
and modifications on electroweak precision observables
arise at  higher orders. Work is in progress in
these directions.

\section*{Acknowledgments}
This work is supported by DGICYT (Spain) under project numbers FPA
2000-0956 and BFM 2002-01003 and by the Fulbright-MEC (Spain)
program. A.D.  acknowledges the hospitality of the SLAC
Theoretical Physics Group and economical support from the
Universidad Complutense del Amo Program.

\end{document}